# Moiré Imaging in Twisted Bilayer Graphene Aligned on Hexagonal Boron Nitride

Xiong Huang[1,2†], Lingxiu Chen[3†], Shujie Tang[4], Chengxin Jiang[4], Chen Chen[4], Huishan Wang[4], Zhi-Xun Shen[5], Haomin Wang[4], Yong-Tao Cui[1*]

[1] Department of Physics and Astronomy, University of California, Riverside, California, 92521, USA
[2] Department of Materials Science and Engineering, University of California, Riverside, California, 92521, USA
[3] School of Materials Science and Physics, China University of Mining and Technology, Xuzhou, 221116, China
[4] State Key Laboratory of Functional Materials for Informatics, Shanghai Institute of Microsystem and Information Technology, Chinese Academy of Sciences, Shanghai 200050, China
[5] Department of Physics and Applied Physics, Geballe Laboratory for Advanced Materials, Stanford University, Stanford, CA 94305, USA

[†] These authors contribute equally to this work.
[*] Correspondence to: yongtao.cui@ucr.edu

**Abstract:** Moiré superlattices (MSL) formed in angle-aligned bilayers of van der Waals materials have become a promising platform to realize novel two-dimensional electronic states. Angle-aligned trilayer structures can form two sets of MSLs which could potentially interfere with each other. In this work, we directly image the moiré patterns in both monolayer graphene aligned on hBN and twisted bilayer graphene aligned on hBN, using combined scanning microwave impedance microscopy and conductive atomic force microscopy. Correlation of the two techniques reveals the contrast mechanism for the achieved ultrahigh spatial resolution (<2 nm). We observe two sets of MSLs with different periodicities in the trilayer stack. The smaller MSL breaks the 6-fold rotational symmetry and exhibits abrupt discontinuities at the boundaries of the larger MSL. Using a rigid atomic-stacking model, we demonstrate that the hBN layer considerably modifies the MSL of twisted bilayer graphene. We further analyze its effect on the reciprocal space spectrum of the dual-moiré system.

Stacking two layers of atomically thin materials of similar structures at a controlled alignment angle can form a moiré superlattice (MSL), whose period could reach 10-100's of nm, much greater than those of individual atomic lattices. The periodic modulation in the MSL can induce drastic changes in the electronic structure of one or both layers, which has become a promising platform to explore novel electronic phases in two dimensions. For example, the MSL in a 0º-aligned monolayer graphene/hBN structure can induce secondary Dirac cones in the electronic band structure of the monolayer graphene, leading to a fractal structure in the Landau level spectrum, known as the Hofstadter's butterfly[1–3]. Twisted bilayer graphene with a magic angle around 1.1º can form flat minibands that host strongly correlated Mott insulator and superconducting states[4–6]. This idea has further been realized in various versions of twisted graphene systems as well as angle-aligned bilayers of transition metal dichalcogenides, in which correlated insulating states, such as Mott insulator and generalized Wigner crystal phases, have been observed[7–24]. When a twisted bilayer graphene is further aligned on an hBN layer, ferromagnetism can be induced in this trilayer structure, and quantized anomalous Hall effect has also been observed, indicating the formation of a Chern insulator phase[25,26]. Multiple MSLs can potentially form in such aligned trilayer structures, contributing extra degree of freedom for the electronic band engineering[27–29]. But studies along this direction have been so far very limited.

Direct visualization of MSL is of great importance in the characterization of its geometric structures. Spatially resolved techniques are ideal tools for this purpose as they can provide direct information in the real space. The large periodicity of the MSL has made it possible to image it with various scanning probe microscopy (SPM) techniques[28,30–37], with the benefits of high throughput and relaxed requirements on sample preparations. Here, we report the observation of a dual-moiré pattern in a CVD-grown twisted bilayer graphene (TBG) that is aligned on an hBN substrate, using both scanning microwave impedance microscopy (MIM) and conductive atomic force microscopy (cAFM). Correlating results from these complementary imaging techniques pinpoints the tip-graphene contact resistance as the mechanism for the ultrahigh spatial resolution (better than 2 nm). In the TBG-hBN structure, two sets of MSLs are observed: a larger MSL (~16 nm) due to the 0º-aligned graphene and hBN, and a smaller MSL (3-6 nm) due to the twisted graphene layers. Interestingly, the smaller MSL is distorted and exhibits abrupt discontinuities at the boundaries of the larger MSL. Using a simple atomic stacking model, these features can be recovered by considering all three layers in the formation of the moiré features. We further identify the signatures of the dual-moiré pattern on the reciprocal-space spectrum of the trilayer structure, which could potentially modify the electronic structure of the twisted bilayer graphene.

First, we demonstrate high-resolution MIM imaging of the moiré superlattice in monolayer graphene aligned on hBN. The schematics of the measurement is shown in Fig. 1a. The moiré sample used in our study is grown by chemical vapor deposition (CVD) method, in which a monolayer graphene grows epitaxially on an exfoliated hBN flake[33] (See Methods for more details). Therefore, the alignment angle is precisely 0º, which is expected to generate a moiré lattice with a periodicity of approximately 14 nm. In the MIM measurement, a shielded cantilever probe is in direct contact with the graphene. A microwave signal (around 3 GHz) is routed to the probe, and the reflected signal is demodulated to generate the MIM signals. Figure 1b plots a typical MIM image for a graphene/hBN MSL. It clearly resolves the honeycomb pattern expected for the MSL, with a spatial resolution better than 2 nm. The honeycomb pattern has noticeable distortions in different regions, which is likely due to local strains in the graphene flake formed during the high-temperature growth process. Fourier transform analysis reveals an average periodicity of ~16 nm, corresponding to a tensile strain of ~0.15% in the monolayer graphene.

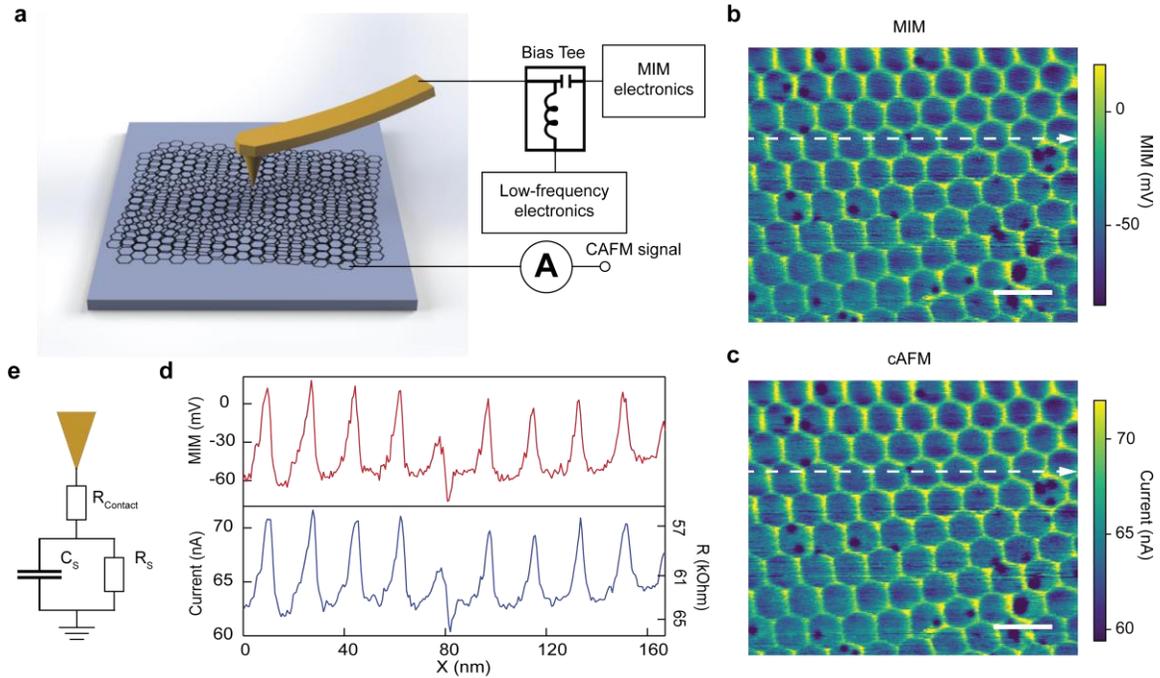

**Figure 1. High resolution imaging of moiré pattern in monolayer graphene/hBN sample. a,** Schematics of the experimental setup. **b & c,** simultaneously acquired (b) MIM and (c) cAFM images. **d,** Linecuts extracted along the dotted lines in MIM and cAFM images. **e,** Effective circuit model for the tip-graphene impedance probed in the contact mode MIM. $R_{contact}$ is the tip-sample contact resistance. $R_S$ nad $C_S$ are the resistance and capacitance of the sample, respectively. Scale bars are 20 nm.

It is worth noting that the ultrahigh spatial resolution is unprecedented in conventional MIM experiments, whose resolution is dominated by the tip radius (typically around 50 nm)[38–40]. Similar results on high-resolution MIM imaging of graphene moiré structures have been reported in two recent experiments[41,42]. However, the contrast mechanism responsible for the high resolution has remain elusive. To gain insights on this issue, we perform cAFM measurement simultaneously with MIM. Using a bias tee, we superpose a low frequency bias voltage on top of the microwave signal and collect the current conducted through the tip-graphene contact using electrical contacts on the graphene flake. In this way, MIM and cAFM images can be acquired simultaneously. Figure 1c plots the cAFM image corresponding to the MIM image in Fig. 1b. These two images show strikingly similar patterns, and even fine features look almost identical as demonstrated in the example linecuts in Fig. 1d. Such high degree of similarity indicates that the contrast mechanism for the MIM signal is the same as that for cAFM, i.e., the tip-graphene contact resistance. As illustrated in Fig. 1e, when the tip is in good contact with the graphene flake, a contact resistance forms at the tip-graphene junction, which allows the low frequency current to flow through. It also contributes to the total tip-sample impedance at microwave frequency, which modulates the microwave reflection and produces the MIM signal contrast. We note that this regime of low tip-sample contact resistance is not readily achieved in typical MIM measurements in which the tip-sample electrical contact is usually dominated by a thin insulating spacer layer. As a result, the typical MIM response curves, obtained by modeling the tip-sample contact as a capacitance, do not apply to this case. (See SI for a comparison of the two effective circuit models.) We would like to point out that, unlike cAFM, the MIM measurement does not require a counter electrode on the sample, making it more versatile to characterize moiré structures without going through extensive fabrication processes for making electrical leads.

Next, we explore the moiré superstructure formed in a trilayer stack in which a second graphene layer (G2) is stacked at an angle relative to the first graphene layer (G1) that has a 0º alignment on the hBN substrate. We expect two sets of moiré patterns, one from 0º aligned G1 and hBN, and the other from G2/G1 with a twist angle $\theta$. We focus on the range $\theta \geq 1.8º$ so that the period of the G2/G1 moiré is much smaller than that of G1/hBN (~14 nm). Figure 2 presents our analysis results based on an atomic stacking model which simply considers the stacking of three rigid layers without any lattice relaxation or reconstruction. We find that the dual-moiré superstructure (Fig. 2c) consists of two sets of periodic patterns – one set with a smaller period (moiré-S) is divided into periodic domains with a larger period (moiré-L). Interestingly, while the two sets resemble the moiré patterns in individual bilayer structures of G2/G1 (Fig. 2b) and G1/hBN (Fig. 2a), it is not a simple combination of the two. Most notably, the moiré-S pattern is along a different direction compared to the G2/G1 pattern, and it also has a discontinuity near the domain boundary between two moiré-L hexagons. We further calculate the three unit vectors for moiré-S. For each unit vector, we extract its magnitude and direction relative to the corresponding unit vector in the G1 graphene layer. Figure 2d and 2e plot their magnitudes and angles as a function of the twist angle $\theta$. We find that, compared to the G2/G1 moiré pattern in which the three unit vectors have equal magnitudes and are uniformly spaced at 60º apart (the black curves in Figs. 2d and 2e), the moiré-S unit vectors have smaller and unequal magnitudes and their orientations are also significantly deviated from those in G2/G1. Such deviations from G2/G1 are more pronounced at small twist angles. These behaviors suggest that the presence of an hBN layer should modify the moiré patterns in a twisted bilayer graphene. Without the hBN layer, the center of each moiré hexagon corresponds to the AA stacking location of the two layers (Fig. 2f). In the presence of the hBN layer, the center now should correspond to a nearly AAA stacking location of three layers where the three A-site atoms are closest to each other (Fig. 2g), which can be different from where the two A-site atoms from G1 and G2 are closest to each other.

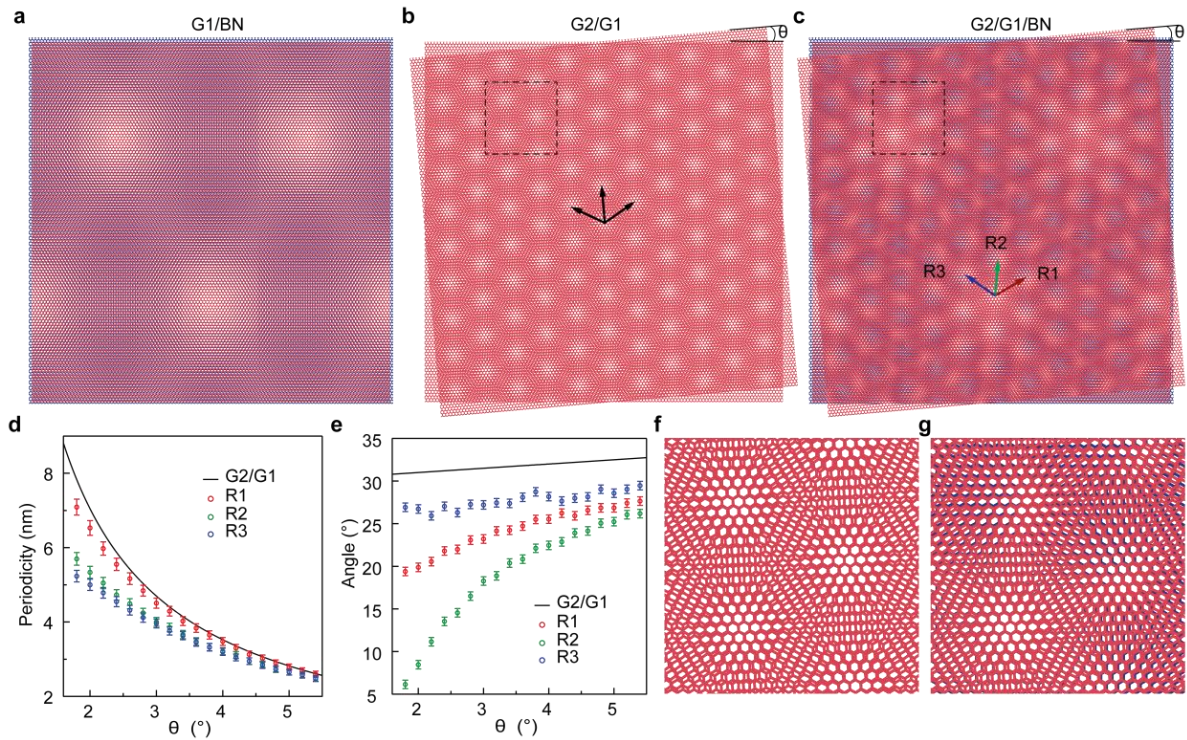

**Figure 2. Analysis of the moiré patterns in a trilayer stacking. a,** 0°-aligned graphene/hBN. **b,** Twisted bilayer graphene with a twist angle $\theta = 5°$. The black arrows correspond to the unit vectors of the moiré lattice. **c,** Twisted bilayer graphene ($\theta = 5°$) on hBN with a 0° angle between G1 and hBN. The arrows labelled by R1-R3 correspond to the unit vectors of the small moiré lattice. **d & e,** The magnitudes and orientations of the unit vectors in (b) and (c). **f & g,** Zoom-in images for the regions indicated by the dotted squares in (b) and (c), respectively.

We examine these model predictions experimentally by imaging the moiré superstructure in a twisted bilayer graphene aligned on hBN substrate. The samples are grown by the same CVD method. During the growth, a second layer of graphene can grow on top of the first graphene layer at an angle that varies at different locations, which allows us to image moiré patterns of the G2/G1/hBN trilayer structure with many different twist angles between G2 and G1. Figures 3a and 3c show two example images taken with MIM (Fig. 3a) or cAFM (Fig. 3c) in the contact mode (images of more samples of different twist angles are presented in SI). In each image, we can identify two periodic patterns of different length scales corresponding to the moiré-S and moiré-L patterns in the model. The moiré-S exhibits a triangular pattern of clear contrast with bright spots corresponding to a low tip-sample contact resistance. According to our analysis, these spots are the AAA stacking regions of the trilayer structure. The moiré-L pattern is marked by boundaries of hexagon domains. These boundaries typically show an enhanced signal (a lower tip-sample contact resistance), and more strikingly, the moiré-S pattern has notable discontinuities at all the moiré-L domain boundaries, exactly as expected from our model. To gain more quantitative comparisons, we analyze the unit vectors of the moiré-S patterns and plot them in a polar plot in Fig. 3d, together with those calculated from our model. The measurement results agree well with model calculations. In particular, three unit vectors have unequal magnitudes and nonuniform angular spacings. These features are most pronounced in the case of Fig. 3a which has a larger period corresponding to a smaller twist angle (Fig. 3b). From Fig. 3d, we further determine the twist angle for all the observed dual-moiré images and the values range from 2° to 5° (see SI). These two prominent features of the moiré-S pattern, i.e., the discontinuities at domain boundaries and the unequal unit vectors, are clear evidences that the aligned hBN substrate modulates the local electronic properties of the twisted bilayer graphene.

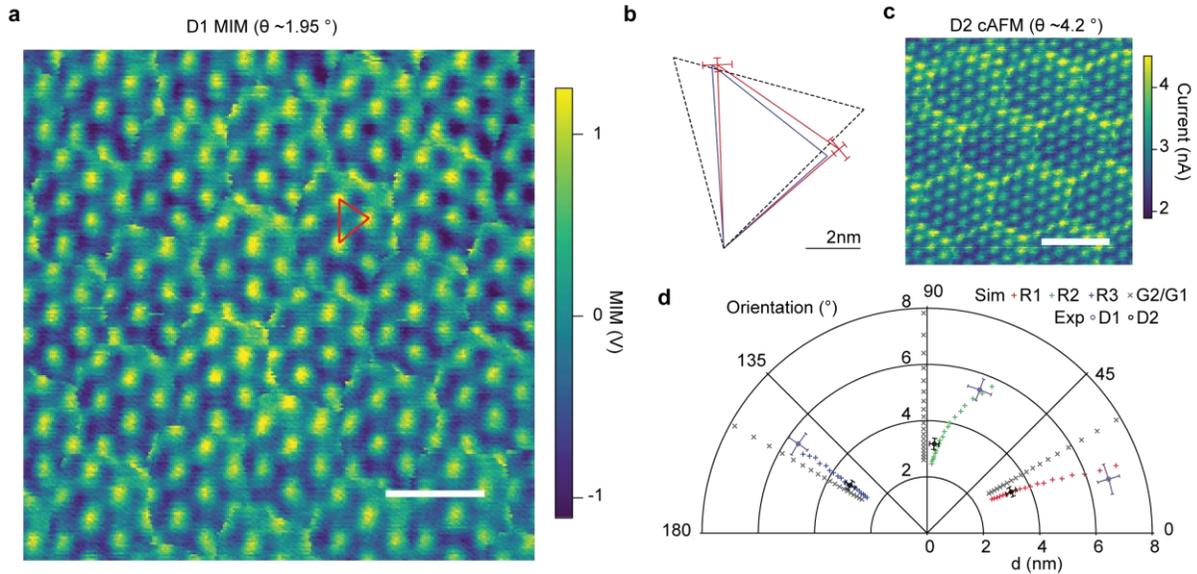

**Figure 3. Imaging dual-moiré pattern in G2/G1/hBN trilayer stacking. a,** MIM image of a trilayer sample D1 with a 1.95° twist angle between G1 and G2. **b,** Comparison of the unit vectors determined from experiment (red) and those from a trilayer stacking model (blue) and a bilayer stacking model (black). The error bars are determined from the uncertainties of the orientations of the moiré-L lattice in the experimental data. **c,** cAFM image of a dual-moiré pattern for a trilayer sample D2 with a 4.2° twist angle between G1 and G2. **d,** Unit vectors of the moiré-S lattice for samples D1 and D2, compared with those calculated from a trilayer stacking model (R1, R2 and R3) and from a bilayer stacking model (G2/G1). Scale bars: (a & c) 15 nm, (b) 2 nm.

Finally, we demonstrate the signatures of such dual-moiré pattern in the reciprocal space, focusing on the effect of the discontinuity at the domain boundaries. We perform MIM imaging over a large area in the same region as the data in Fig. 3a. The MIM image (Fig. 4a) shows that the dual moiré pattern fills the entire area with only few local variations in the signal intensity. We then perform fast Fourier transform (FFT) of the MIM image and plot the spectrum in Fig. 4b. The spectrum contains a series of peaks which can be identified in the following categories. A dashed hexagon centered around the origin corresponds to the moiré-L pattern. Bright spots at large $k$ values correspond to the features from the moiré-S pattern. Since the moiré-S pattern is spatially modulated by the moiré-L boundaries, the moiré-L hexagon is further replicated at each of the bright spots in the moiré-S pattern. Other peaks are the higher order peaks of these replica. To better illustrate their relation, we sketch only the main features in Fig. 4c. We note that the moiré-S pattern contains the discontinuities at the moiré-L boundaries. To examine the effect of such discontinuities, we plot the reciprocal vectors ($k'_{moiré-S}$, labeled with dashed purple arrows) calculated directly from the unit vectors identified in Fig. 3a. These $k$ vectors essentially represent the FFT peaks for a continuous moiré-S pattern without any modulation by the moiré-L pattern.

Their positions significantly deviate from the experimentally obtained peak positions ($k_{\text{moiré-S}}$, labeled with solid purple arrows), and the differences are caused by the discontinuities of the moiré-S pattern across the moiré-L domain boundaries, which are well described by our analytical calculations presented in the Supplementary Information.

Furthermore, we also plot the $k_{\text{GG}}$ vectors (the gray arrows) corresponding to a twisted bilayer graphene with a twist angle of 1.95°, the same as that determined in the trilayer structure measured in Fig. 4a. From the comparison of $k_{\text{GG}}$ with $k_{\text{moiré-S}}$ and $k'_{\text{moiré-S}}$, we identify the effect in the $k$-space structure of the twisted bilayer graphene due to the aligned hBN layer. As discussed in Fig. 2, the aligned hBN layer changes the local moiré-S pattern by a small rotation as well as a distortion that breaks the original 6-fold rotational symmetry, which is reflected in the $k$-space as the change from $k_{\text{GG}}$ to $k'_{\text{moiré-S}}$. Then the discontinuities at the moiré-L boundaries further shift $k'_{\text{moiré-S}}$ to $k_{\text{moiré-S}}$. The $k$-space structure probed in our study reflects the spatial modulation of the tip-sample contact resistance which originates from spatial modulation in the electronic structure of the twisted bilayer graphene. Therefore, as revealed by our analysis, the aligned hBN layer is expected to significantly modify the electronic structure of the twisted bilayer graphene in the momentum space. We believe our results could inspire further experimental and theoretical studies, such as band structure calculations, along these directions.

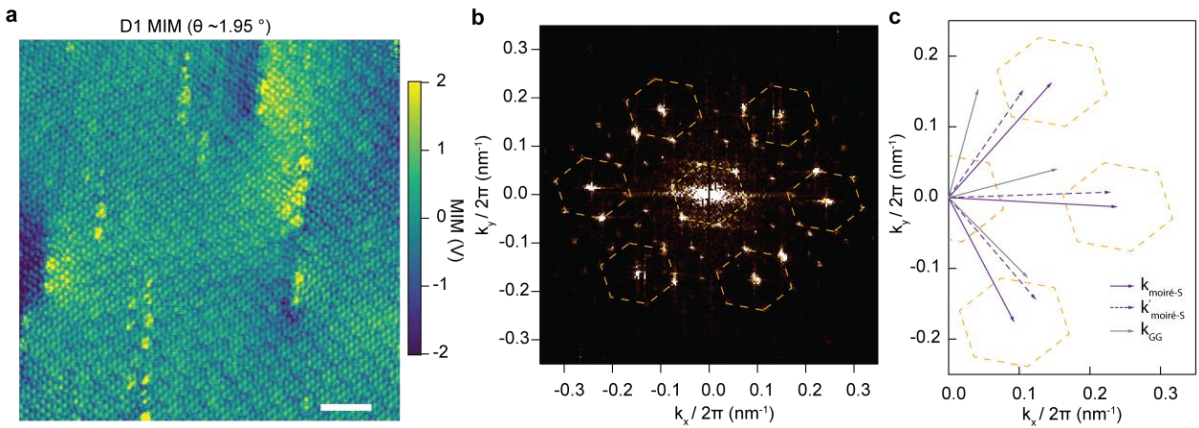

**Figure 4. Analysis of the dual-moiré pattern in the reciprocal space. a,** MIM image of a large area in sample D1. Scale bar: 40 nm. **b,** Fourier transform of the MIM image in (a). The dotted hexagon centered at the origin corresponds to the mookiré-L lattice, and the other hexagons are replicas centered at the peaks of the moiré-S lattice. **c,** Comparison of the Fourier peaks of the moiré-S lattice extracted from (b) (solid purple arrows), the reciprocal vectors calculated from the moiré-S unit vectors determined from the real space data in D1 (Fig. 3a) (dotted purple arrows), and the reciprocal vectors calculated for a twisted bilayer graphene with the same twist angle of 1.95° as D1 (gray arrows).

**Methods:**
**Sample preparations:** The graphene layers are deposited on the hBN flakes which are pre-exfoliated on quartz via chemical vapor deposition.[33,43] The hBN flakes are mechanically exfoliated onto quartz and annealed at 600°C in oxygen flow to remove residues. Then the quartz substrate with hBN is placed into a graphite tube and put into the CVD chamber. The growth of twisted bilayer graphene is carried out at 1300°C. First, a mixture of ethyne and silane in pulsed flow is introduced to form the first graphene layer in polycrystalline, and then the mixture is tuned to be in a stable flow to grow the second graphene layer. After that, the substrate cools down slowly to room temperature. For the MIM measurement, the sample requires little preparation. For the cAFM measurement, an electrical contact is made to the polycrystalline graphene layer that is continuous in a large area, using silver epoxy in remote locations far away from tip scan regions.

**MIM and cAFM measurements:** The MIM and cAFM measurements are performed on commercial atomic force microscopes (Bruker Dimension Icon AFM and Asylum Research Cypher AFM). Shielded probes with solid metal probe tip (ScanWave probes from PrimeNano with a spring constant of about 1 N/m) and ASYELEC.01-R2 probes from AsylumResearch (spring constant of about 2.8 N/m) have been used in contact mode scans. A microwave excitation (about 0.01 mW at 3 GHz) and an AC voltage (5 mV at ~7.4 kHz) or DC voltage (5 - 10 mV) are coupled through a bias tee and applied to the cantilever probe. The current flowing through tip-sample junction is amplified and then demodulated using a SR830 lock-in (only for AC voltage bias). The reflected microwave signal is analyzed to extract the demodulated outputs as MIM signals.

**Author Contributions**

X.H. and L.C. contribute equally to this work. H.W., Z.X.S., and Y.T.C. initiated the research. L.C. performed the sample growth with assistance from C.J., C.C., and H.S.W.. X.H. performed the MIM measurement. X.H., L.C., and C.C. performed the cAFM measurements. S.T. carried out initial MIM measurements on graphene/hBN structure. Y.T.C. and X.H. analyzed the data and wrote the manuscript with inputs from all authors.

**Notes**

Z.-X.S. is a cofounder of PrimeNano Inc., which licensed the MIM technology from Stanford for commercial instrument.

**Acknowledgments**

We thank Q. Li for the help on cAFM measurements. X.H. and Y.T.C. acknowledge support from NSF under award DMR-2004701, the Hellman Fellowship award, and the seed fund from SHINES, an EFRC funded by the U. S. Department of Energy (DOE), Basic Energy Sciences (BES) under award number SC0012670. The sample growth and fabrication work was supported by the National Key R&D program (Grant No. 2017YFF0206106), the Strategic Priority Research Program of Chinese Academy of Sciences (Grant No. XDB30000000), National Natural Science Foundation of China (Grant No. 51772317, 91964102, 12004406), the Science and Technology Commission of Shanghai Municipality (Grant No. 20DZ2203600) and Soft Matter Nanofab (SMN180827) of ShanghaiTech University. Z.X.S. acknowledges support from the Gordon and Betty Moore Foundation through Grant No. GBMF4546.

**Data Availability:**

The datasets generated and/or analyzed during this study are available from the corresponding author upon reasonable request.

# Supplementary Information

**S1. MIM effective circuit models and response curves**

**S2. Data on additional samples**

**S3. Atomic stacking model for dual moiré structures**

**S4. FFT analysis of dual-moiré patterns**

## S1. MIM effective circuit models and response curves

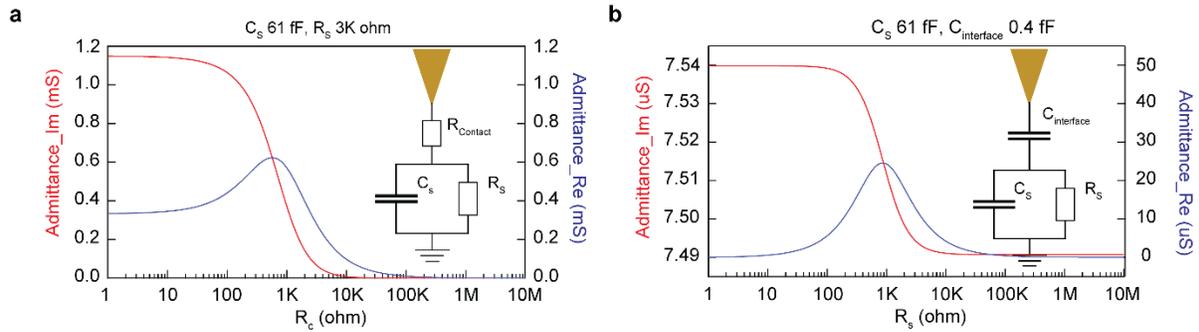

Figure S1. Simulated MIM response curves based on two effective circuit models for the tip-sample interaction. (a) The tip-sample contact has a good electrical conduction which can be modelled by a contact resistance. This is the case for imaging moiré patterns in graphene/hBN samples in this work. The response curves plot the imaginary and real parts of the tip-sample admittance (inverse of impedance) as a function of varying tip-sample contact resistance. (b) The tip-sample contact does not have a good electrical conduction and is thus dominated by an interface capacitance. This is the case for MIM measurements on most of common samples. The response curves plot the imaginary and real parts of the tip-sample admittance as a function of varying sample resistance. Note the different range for the MIM signal span. The case in (a) produces a much larger MIM response, consistent with experimental observations.

## S2. Data on additional samples

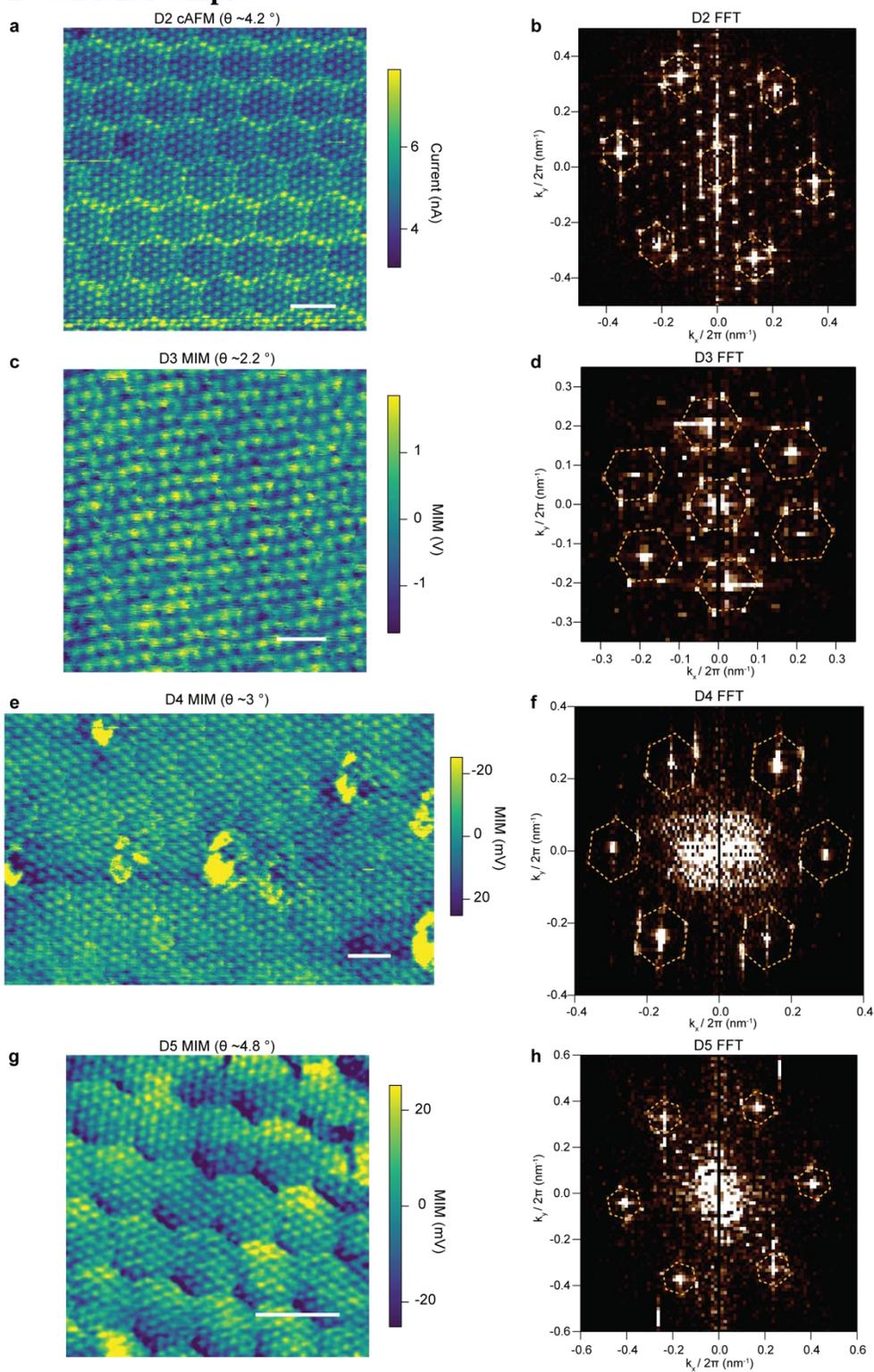

Figure S2. MIM/cAFM images for TBG/hBN samples D2-D5 (a, c, e and g) and their FFT images (b, d, f and h). All scale bars: 15 nm.

## S3. Atomic stacking model for dual-moiré structures

The periodic atomic lattice for each of the three layers in a TBG/hBN structure can be described by the lowest harmonics:

$$f(\vec{r}) = \sum_{i=1}^{3} e^{i\vec{k}_i \cdot \vec{r}}$$

where $\vec{k}_i$ ($i$=1,2,3) are the unit vectors for the lattice. Thus $\vec{k}_i \cdot \vec{r}$ represent the phase of the lattice. The formation of moiré patterns depends on the alignment among the phases of each layer. The center of each moiré-S domain can be identified by the location where the three lattice phases are closest to each other (See an example in Fig. S3a). We thus define a phase variation as

$$\Delta_{phase}(\vec{r}) = (\text{wrap}(\vec{k}_1^{hBN} \cdot \vec{r} - \vec{k}_1^{G1} \cdot \vec{r}))^2 + (\text{wrap}(\vec{k}_2^{hBN} \cdot \vec{r} - \vec{k}_2^{G1} \cdot \vec{r}))^2$$
$$+ (\text{wrap}(\vec{k}_1^{hBN} \cdot \vec{r} - \vec{k}_1^{G2} \cdot \vec{r}))^2 + (\text{wrap}(\vec{k}_2^{hBN} \cdot \vec{r} - \vec{k}_2^{G2} \cdot \vec{r}))^2$$
$$+ (\text{wrap}(\vec{k}_1^{G1} \cdot \vec{r} - \vec{k}_1^{G2} \cdot \vec{r}))^2 + (\text{wrap}(\vec{k}_2^{G1} \cdot \vec{r} - \vec{k}_2^{G2} \cdot \vec{r}))^2$$

Here the wrap() function wraps the phase value to the range $[-\pi, \pi]$. The centers of moiré-S domains correspond to the minimum values of $\Delta_{phase}$. Fig. S3b plots the calculated $\Delta_{phase}$ for the case in Fig. S3a. The markers indicate the locations of the minimum values of $\Delta_{phase}$, which match well the atomic patterns in Fig. S3a. The unit vectors for the moiré-S pattern can be extracted from these locations. This model allows us to efficiently calculate the lattice parameters for the moiré-S patterns in the TBG/hBN structure with a varying twist angle between the two graphene layers.

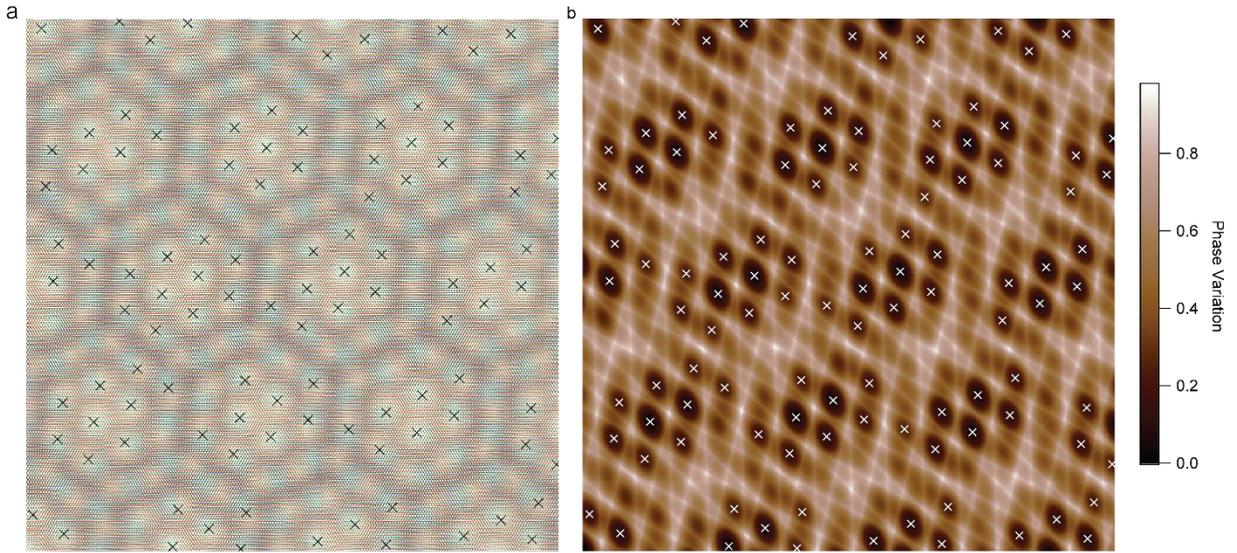

Figure S3. (a) Atomic stacking model for twisted bilayer graphene (3.6°) on hBN. Blue atoms for hBN, red atoms for aligned graphene and green atoms for rotated graphene, respectively. (b) Calculated phase variation $\Delta_{phase}$. Locations of the minimum values of $\Delta_{phase}$ are indicated by the markers in both (a) and (b).

## S4. FFT analysis on dual-moiré patterns

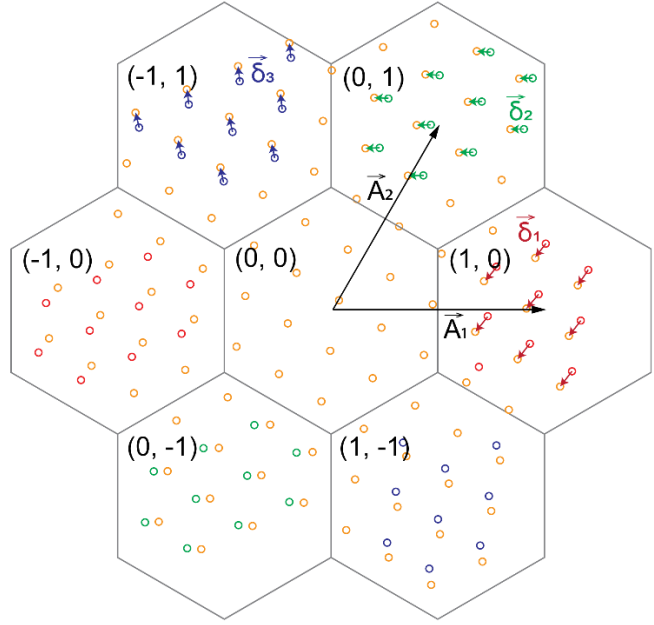

Figure S4. FFT analysis of a periodic pattern with discontinuities at the boundaries of the large periodic structure. Orange circles are the lattice extended from domain (0, 0) without including any discontinuities. Red circles are those for hexagon domains (1, 0) and (-1, 0). Green (blue) ones are for domains (0, 1) and (0, -1) ((-1, 1) and (1, -1)).

In this section, we present the mathematical analysis on the Fourier transform of a dual moiré superstructure in which the moiré-S pattern has discontinuities at the boundaries of the moiré-L unit cell domains.

The periodic structure of the triangular moiré-S lattice can be described by the lowest harmonics of the following form:

$$f(\vec{r}) = \sum_{i=1}^{3} e^{i\vec{k}_i \cdot \vec{r}}$$

with $\vec{k}_1$, $\vec{k}_2$ and $\vec{k}_3 = \vec{k}_2 - \vec{k}_1$ the three unit vectors of the moiré-S lattice.

Across the boundaries of the moiré-L domains, the discontinuities in the moiré-S lattice can be modelled as two displacement vectors, $\vec{\delta}_1$ and $\vec{\delta}_2$, along the two directions defined by the two unit vectors of the moiré-L lattice, $\vec{A}_1$ and $\vec{A}_2$, respectively. In the (m, n)-th unit cell of the moiré-L lattice, as shown in Fig. S4, the overall accumulated displacement will be:

$$\vec{\delta}_{mn} = m\vec{\delta}_1 + n\vec{\delta}_2$$

Therefore, in general, the moiré-S lattice with discontinuities can be expressed as

$$f(\vec{r}) = \sum_{i=1}^{3} e^{i\vec{k}_i \cdot (\vec{r} - \vec{\delta}_{mn})}$$

for $\vec{r}$ within the (m, n)-th domain. Therefore, the Fourier transform of $f(\vec{r})$ can be obtained by calculating the Fourier transform of each term, $F_i(\vec{k})$. Thus, we have:

$$F_i(\vec{k}) = \iint e^{i\vec{k}_i \cdot (\vec{r} - \vec{\delta}_{mn})} \cdot e^{-i\vec{k} \cdot \vec{r}} d^2\vec{r}$$

where the integral is over the entire plane.

We evaluate the integral by summing the contribution from individual unit cells. Specifically, in the (m, n)-th unit cell which is centered at $\vec{R}_{m,n} = m \cdot \vec{A}_1 + n \cdot \vec{A}_2$, we define a reduced position vector $\vec{r}_0 = \vec{r} - \vec{R}_{m,n}$, and we express the integral in terms of $\vec{r}_0$.

$$F_i(\vec{k}) = \sum_{m,n} F_i^{mn}(\vec{k}) = \sum_{m,n} \iint_{u.c.} e^{i\vec{k}_i \cdot (\vec{r}_0 + \vec{R}_{m,n} - m\cdot\vec{\delta}_1 - n\cdot\vec{\delta}_2)} \cdot e^{-i\vec{k}\cdot(\vec{r}_0 + \vec{R}_{m,n})} d^2\vec{r}_0$$

$$= \sum_{m,n} e^{i\vec{k}_i \cdot (\vec{R}_{m,n} - m\cdot\vec{\delta}_1 - n\cdot\vec{\delta}_2) - i\vec{k}\cdot\vec{R}_{m,n}} \iint_{u.c.} e^{i(\vec{k}_i - \vec{k})\cdot\vec{r}_0} d^2\vec{r}_0$$

$$= \sum_{m,n} e^{i\vec{k}_i \cdot (\vec{R}_{m,n} - m\cdot\vec{\delta}_1 - n\cdot\vec{\delta}_2) - i\vec{k}\cdot\vec{R}_{m,n}} F_i^{00}(\vec{k})$$

Note that $F_i^{00}(\vec{k}) \equiv \iint_{u.c.} e^{i(\vec{k}_i - \vec{k})\cdot\vec{r}_0} d^2\vec{r}_0$ is simply the integral within the (0, 0)-th unit cell which is a constant independent of $m$ and $n$. We then have

$$F_i(\vec{k}) = F_i^{00}(\vec{k}) \sum_{m,n} e^{i\vec{k}_i \cdot (\vec{R}_{m,n} - m\cdot\vec{\delta}_1 - n\cdot\vec{\delta}_2) - i\vec{k}\cdot\vec{R}_{m,n}}$$

$$= F_i^{00}(\vec{k}) \sum_{m,n} e^{i[(\vec{k}_i - \vec{k})\cdot(m\cdot\vec{A}_1 + n\cdot\vec{A}_2) - m\vec{k}_i\cdot\vec{\delta}_1 - n\vec{k}_i\cdot\vec{\delta}_2]}$$

$$= F_i^{00}(\vec{k}) \sum_{m} e^{im\cdot\left((\vec{k}_i - \vec{k})\cdot\vec{A}_1 - \vec{k}_i\cdot\vec{\delta}_1\right)} \sum_{n} e^{in\cdot\left((\vec{k}_i - \vec{k})\cdot\vec{A}_2 - \vec{k}_i\cdot\vec{\delta}_2\right)}$$

$$= F_i^{00}(\vec{k}) \frac{1 - e^{iM[(\vec{k}_i - \vec{k})\cdot\vec{A}_1 - \vec{k}_i\cdot\vec{\delta}_1]}}{1 - e^{i[(\vec{k}_i - \vec{k})\cdot\vec{A}_1 - \vec{k}_i\cdot\vec{\delta}_1]}} \frac{1 - e^{iN[(\vec{k}_i - \vec{k})\cdot\vec{A}_2 - \vec{k}_i\cdot\vec{\delta}_2]}}{1 - e^{i[(\vec{k}_i - \vec{k})\cdot\vec{A}_2 - \vec{k}_i\cdot\vec{\delta}_2]}}$$

$$|F_i(\vec{k})| = \frac{\sin\left(\frac{M}{2}[(\vec{k}_i - \vec{k})\cdot\vec{A}_1 - \vec{k}_i\cdot\vec{\delta}_1]\right)}{\sin\left(\frac{1}{2}[(\vec{k}_i - \vec{k})\cdot\vec{A}_1 - \vec{k}_i\cdot\vec{\delta}_1]\right)} \cdot \frac{\sin\left(\frac{N}{2}[(\vec{k}_i - \vec{k})\cdot\vec{A}_2 - \vec{k}_i\cdot\vec{\delta}_2]\right)}{\sin\left(\frac{1}{2}[(\vec{k}_i - \vec{k})\cdot\vec{A}_2 - \vec{k}_i\cdot\vec{\delta}_2]\right)} F_i^{00}(\vec{k})$$

The peak positions of $|F_i(\vec{k})|$, denoted as $\vec{k}_i'$, satisfy the conditions:

$$(\vec{k}_i - \vec{k}_i') \cdot \vec{A}_1 - \vec{k}_i \cdot \vec{\delta}_1 = 0 \quad \& \quad (\vec{k}_i - \vec{k}_i') \cdot \vec{A}_2 - \vec{k}_i \cdot \vec{\delta}_2 = 0$$

Define the reciprocal vectors $\vec{k}_{A1}$ and $\vec{k}_{A2}$ as

$$\vec{k}_{A1} = \frac{\vec{A}_2 \times \vec{e}_z}{|\vec{A}_1 \times \vec{A}_2|}$$

$$\vec{k}_{A2} = \frac{\vec{e}_z \times \vec{A}_1}{|\vec{A}_1 \times \vec{A}_2|}$$

And they satisfy the relations $\vec{k}_{A1} \cdot \vec{A}_1 = 1$ and $\vec{k}_{A2} \cdot \vec{A}_2 = 1$. We can then obtain

$$\vec{k}_i - \vec{k}'_i = \vec{k}_{A1} \cdot (\vec{k}_i \cdot \vec{\delta}_1) + \vec{k}_{A2} \cdot (\vec{k}_i \cdot \vec{\delta}_2)$$

or

$$\vec{k}'_i = \vec{k}_i - [\vec{k}_{A1} \cdot (\vec{k}_i \cdot \vec{\delta}_1) - \vec{k}_{A2} \cdot (\vec{k}_i \cdot \vec{\delta}_2)]$$

We can see that the Fourier peak positions of the discontinuous lattice, $\vec{k}'_i$, are shifted from those for a continuous lattice, $\vec{k}_i$, by a vector that depends on the displacement vectors, $\vec{\delta}_1$ and $\vec{\delta}_2$, and the reciprocal vectors corresponding to the moiré-L lattice, $\vec{k}_{A1}$ and $\vec{k}_{A2}$.

Experimentally, we can find $\vec{k}'_i$ from the Fourier transform of the entire image containing large number of moiré-L unit cells, and $\vec{k}_i$ can be determined from the Fourier transform of the data within individual moiré-L unit cells. $\vec{k}_{A1}$ and $\vec{k}_{A2}$ can also be determined by extracting the unit vectors for the moiré-L lattice. From these, we can then determine the displacement vectors, $\vec{\delta}_1$ and $\vec{\delta}_2$. Alternatively, $\vec{\delta}_1$ and $\vec{\delta}_2$ can also be determined by comparing the moiré-S lattice patterns in neighboring moiré-L unit cells. The results from these two methods match very well.